\documentstyle[epsf]{mn} 
\begin{document}
\title[Late-type variables in Baade's windows]
{AGB variables in Baade's windows} \author[M. Schultheis and I.S. 
Glass] {M. Schultheis$^1$, I.S. Glass$^2$ 
\\ $^1$Institut d'Astrophysique, 98 bis Blvd Arago, F75014 Paris, France (email: schulthe@iap.fr)\\
$^2$South African Astronomical Observatory, PO Box 9, Observatory 7935, South 
Africa (email: isg@saao.ac.za)}

\date{Received 2000}

\maketitle

\begin{abstract}
In this work, a sample of luminous M-type giants in the Baade's Windows
towards the inner Galactic Bulge is investigated in the near-infrared. The
ISOGAL survey at 7 and 15\,$\mu$m has given information concerning the
mass-loss rates of these stars and their variability characteristics have
been extracted from the MACHO database. Most are known to be semi-regular
variables (SRVs).  Here we discuss how their $IJHK_S-$region colours depend
on period and the presence or absence of mass-loss, using results mainly
taken from the DENIS and 2MASS surveys.

In order to compare their colours with solar neighbourhood stars,
photometric colours on the DENIS, 2MASS and ESO photometric systems have
been synthesized for objects in the spectrophotometric atlas of Lan\c{c}on
and Wood (2000). In addition, they have been used to predict the differences
in colour indicies when stars with strong molecular bands are observed using
different photometric systems.

The SRVs are found to inhabit the upper end of the $J-K_S$, $K_S$
colour-magnitude diagram, lying just below the Miras. High mass-loss rates
are associated with high luminosity. The near-infrared colours of the
semi-regular variables increase in a general way with period and are reddest
for the stars with significant mass-loss. The average colours of Mira
variables, whose periods start at around 200 days in the Bulge, are bluer
than those of the semi-regulars at this period, particularly in $J-H$,
thanks to the association of deep water-vapour bands with large amplitude.

\end{abstract}


\section{Introduction}

The ISOGAL programme (Omont et al. 1999, 2001) has made use of the ISO
satellite to obtain detailed surveys of sample areas of the inner Bulge and
the Galactic Plane at two mid-infrared wavelengths, 7$\mu$m and 15$\mu$m,
with the intention of investigating the longitude and latitude variations of
stellar properties and their bearing on galactic structure. Because many of
the fields are highly obscured, parts of the relatively clear Baade's
Windows were included for comparison purposes (Glass et al. 1999). By
studying the contents of the latter at infrared and visible wavelengths one
obtains a picture of the basic stellar population in the Bulge, to which
more heavily obscured fields, observable only in the infrared, can be
compared. Fields of 15 $\times$ 15 arcmin$^2$ from each of Sgr I and
NGC\,6522 were thus included in the ISOGAL work. These were centred at
$\ell$ = +1.37$^{\circ}$, $b$ = --2.63$^{\circ}$ and $\ell$ =
+1.03$^{\circ}$, $b$ = --3.83$^{\circ}$ respectively.

Near-infrared data at $I$(0.79\,$\mu$m), $J$(1.22\,$\mu$m) and
$K_S$(2.14\,$\mu$m), where the subscript $S$ denotes `short', for both
windows are now available from the DENIS survey (Epchtein, 1998) and for the
Sgr\,I window only at $J$($\sim$1.22\,$\mu$m), $H$(1.65\,$\mu$m) and
$K_S$(2.16\,$\mu$m) from 2MASS (Skrutskie, 1998). In addition, photometry on
the Caltech-CTIO system by Frogel and Whitford (1987, hereafter FW) was
obtained for many of the M stars in the objective prism survey of the
NGC\,6522 field by Blanco, McCarthy and Blanco (1984) and Blanco (1986). A
comparison of stars detected by ISOGAL with the latter survey indicates
that, except for a few probable foreground stars, none with spectral type
earlier than M2 were seen. The rate of detection increased with sub-type,
reaching 100\% at M6.

Several previous surveys of these fields for variable stars have been made.
$I$-band photography by Lloyd Evans (1976) yielded identifications and
periods of many Miras. A search for the near-infrared counterparts of IRAS
sources (Glass 1986) revealed some additional long-period, large-amplitude
variables that were not found in the $I$-band survey. Infrared observations
of nearly all the long-period, large-amplitude variables in Sgr I were
obtained by Glass et al. (1995), so that the sample is probably complete and
the period distribution, period-luminosity relation and period-colour
properties of Miras in Sgr I may be regarded as known.

The Sgr\,I and NGC\,6522 windows formed part of the MACHO gravitational lens
survey. Alard et al. (2001) present variability information for all 332
stars that have been detected at both ISOGAL and both MACHO wavelengths:
$v$\,($\sim$0.54$\mu$m), $r$\,($\sim$0.70$\mu$m), 7\,$\mu$m and 15\,$\mu$m.
Nearly all of the 332 objects were found to be semi-regular variables (SRVs)
with periods of from 10 to 230 days and amplitudes $<$ 1 mag.  and will be
referred to collectively here as the ISOGAL/MACHO sample. The SRVs in the
present fields outnumber the Miras by a factor of about 20. Note that the
area surveyed by MACHO omits about 10\% of each ISOGAL field.

%It is probable
%that, if the ISOGAL survey had been continued to lower flux levels, more
%variables with small ($<$ 0.2 mag) amplitudes would have been found (see,
%e.g., Eyer \& Grenon, 1997; Koen \& Laney, 2000).

We have assumed that the interstellar absorption is $A_V$=1.5 mag for both
fields. We take in addition $A_I$=0.59$A_V$, $A_J$=0.245$A_V$,
$A_H$=0.142$A_V$ and $A_K$=0.085$A_V$, based on the van de Hulst curve
(Glass, 1999).

\section{Comparison of DENIS, 2MASS and other photometry}

All objects falling in the ISOGAL fields have been extracted from the DENIS
and 2MASS databases as well as the FW work. A search radius of 2$''$ was
chosen to avoid mis-identifications.  The DENIS observations form part of a
dedicated survey of the Galactic Bulge (Simon et al., in preparation). The
2MASS observations are limited to Sgr\,I and a small part of NGC\,6522 with
the consequence that only 225 of the 332 ISOGAL/MACHO stars fall within the
common area. In this and the following section we show by direct comparisons
that the available information is consistent.

Firstly, it is necessary to consider what effect the different filter
systems may be expected to have on the photometry. The DENIS filters have been convolved with the typical atmospheric
transmission at La Silla (Fouqu\'{e} et al. 2000), where the survey was
made. A summary of the available data is given in table 1. For details of
the filter transmissions and systematic response curves, see Fouqu\'{e} et
al (2000) for DENIS; 2MASS web pages for 2MASS.

\begin{table}
\caption{Filter transmission data}
\begin{tabular}{lll}
DENIS\\
Band  & cut-on ($\mu$m) & cut-off ($\mu$m)\\
$I$   & 0.73 & 0.87 \\
$J$   & 1.10 & 1.39 \\
$K_S$ & 1.98 & 2.31 \\
\\
2MASS\\
Band & cut-on ($\mu$m) & cut-off ($\mu$m)\\
$J$   & 1.11 & 1.36 \\
$H$   & 1.50 & 1.80 \\
$K_S$ & 2.00 & 2.32 \\
\\
\end{tabular}

Notes:\\ 

The cut-on and cut off wavelengths are taken to be the 50\% transmission
values.

The DENIS transmissions include effects due to the earth's atmosphere. The
effective long wavelength end of the $J$-band is determined by atmospheric
transmission at about 1.35$\mu$m.

\end{table}

\subsection{Photometric passbands}

Important differences between systems occur in the $J$- and $K$-bands. The
Caltech-CTIO system as used by FW  has an effective $J$ wavelength
(see $J_{\rm old}$ in fig.\ 4 of Persson et al. 1998) of about 1.25$\mu$m,
considerably longer than that of DENIS ($\sim$1.22\,$\mu$m).

\begin{figure}
%fig1
\epsfxsize=8.3cm
\epsffile[20 17 592 779]{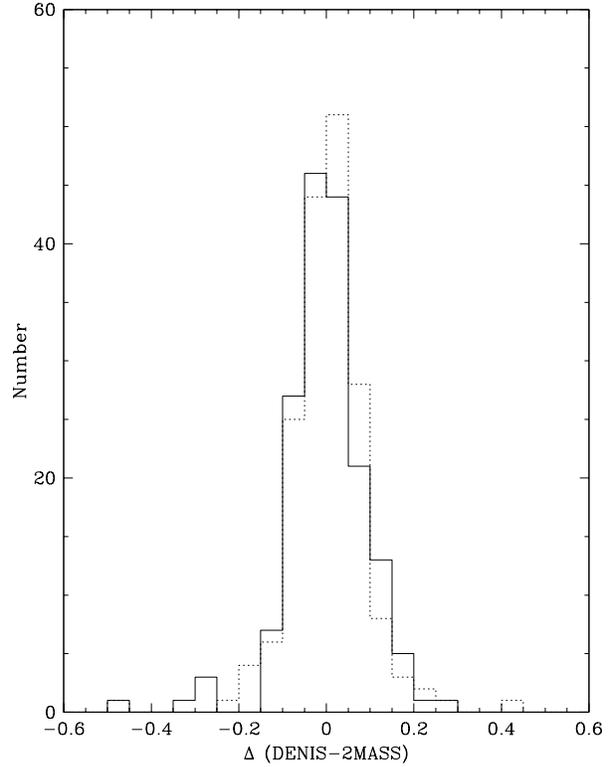}
\caption{Histograms of differences between DENIS and 2MASS $J$ and $K_S$ data
for ISOGAL/MACHO variables identified in both data sets. The solid line is $J$ and
the dotted line is $K_S$.}
\end{figure}

The $H$ and $K$ bands of the SAAO photometric system are very close to those
of the ESO system (see Bouchet, Schmider \& Manfroid, 1991). The SAAO Mira
data for the Baade's Window field (Glass et al. 1995), to which we refer
later on, were taken with the MkIII photometer, whose effective wavelength
at $J$ is $\sim$1.22\,$\mu$m (Glass, 1993). This is essentially identical to
$J_{\rm DENIS}$.

\begin{figure*} 
%fig2
{\epsfxsize=5.7cm \epsfbox[20 17 592 779]{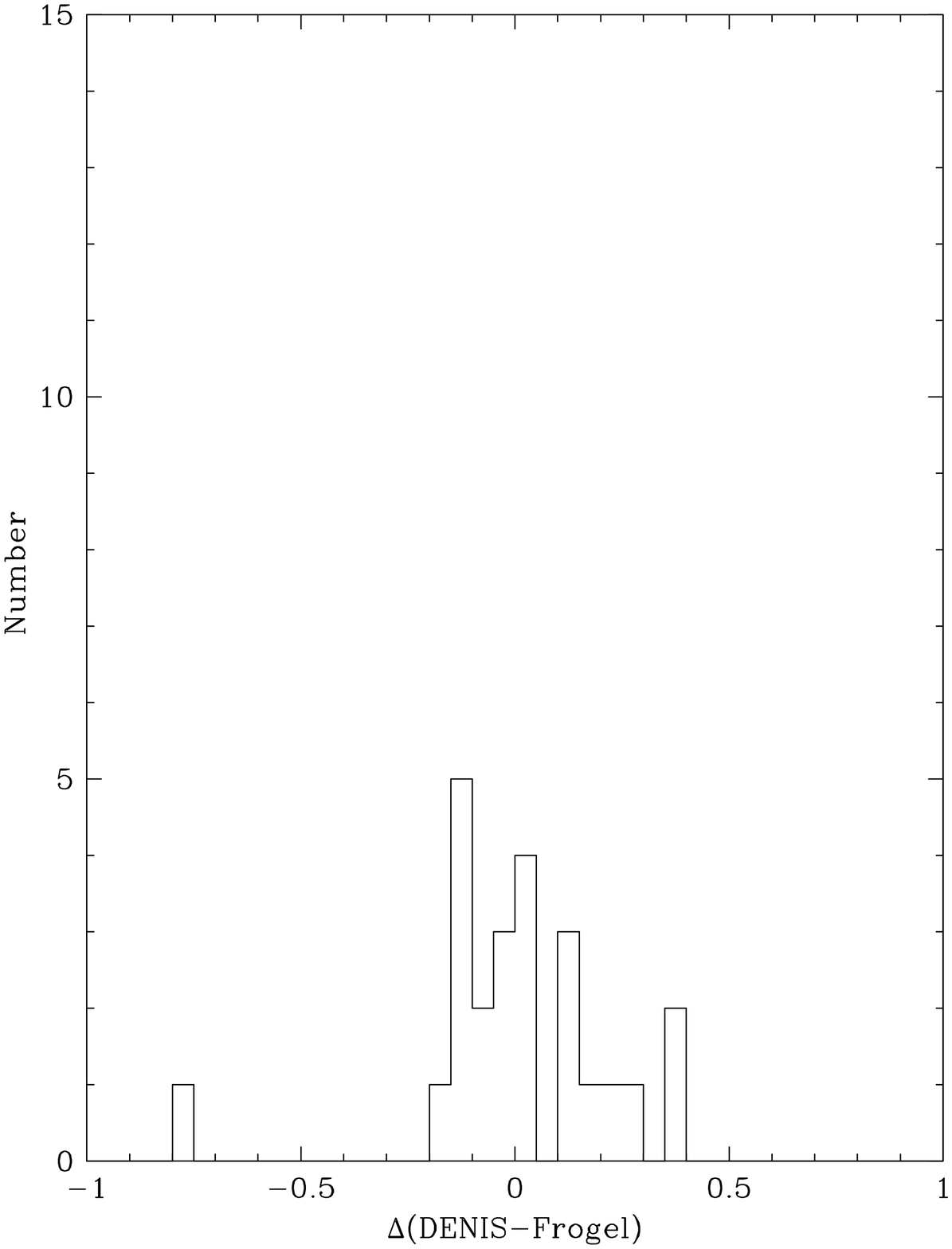}
\epsfxsize=5.7cm \epsfbox[20 17 592 779]{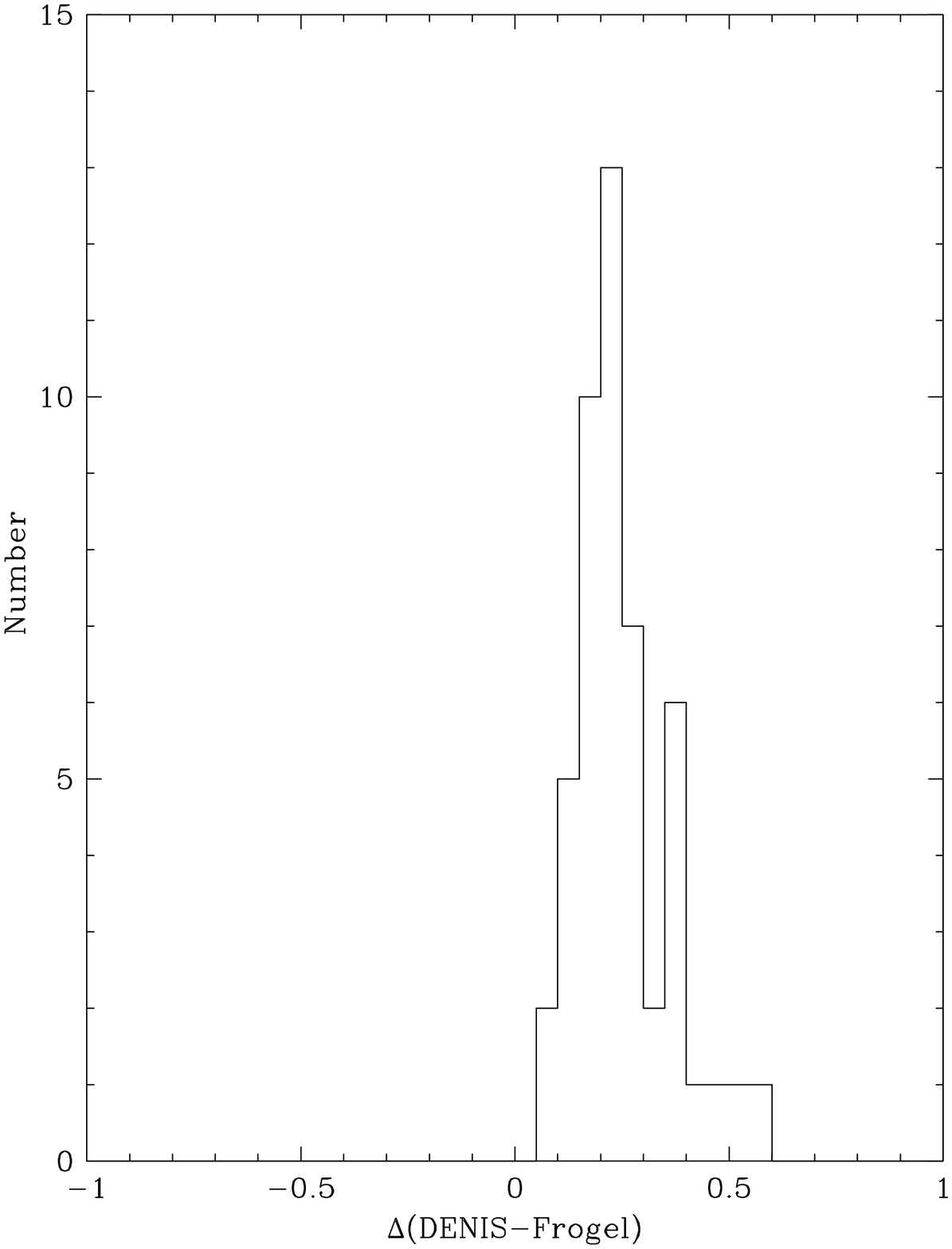}
\epsfxsize=5.7cm \epsfbox[20 17 592 779]{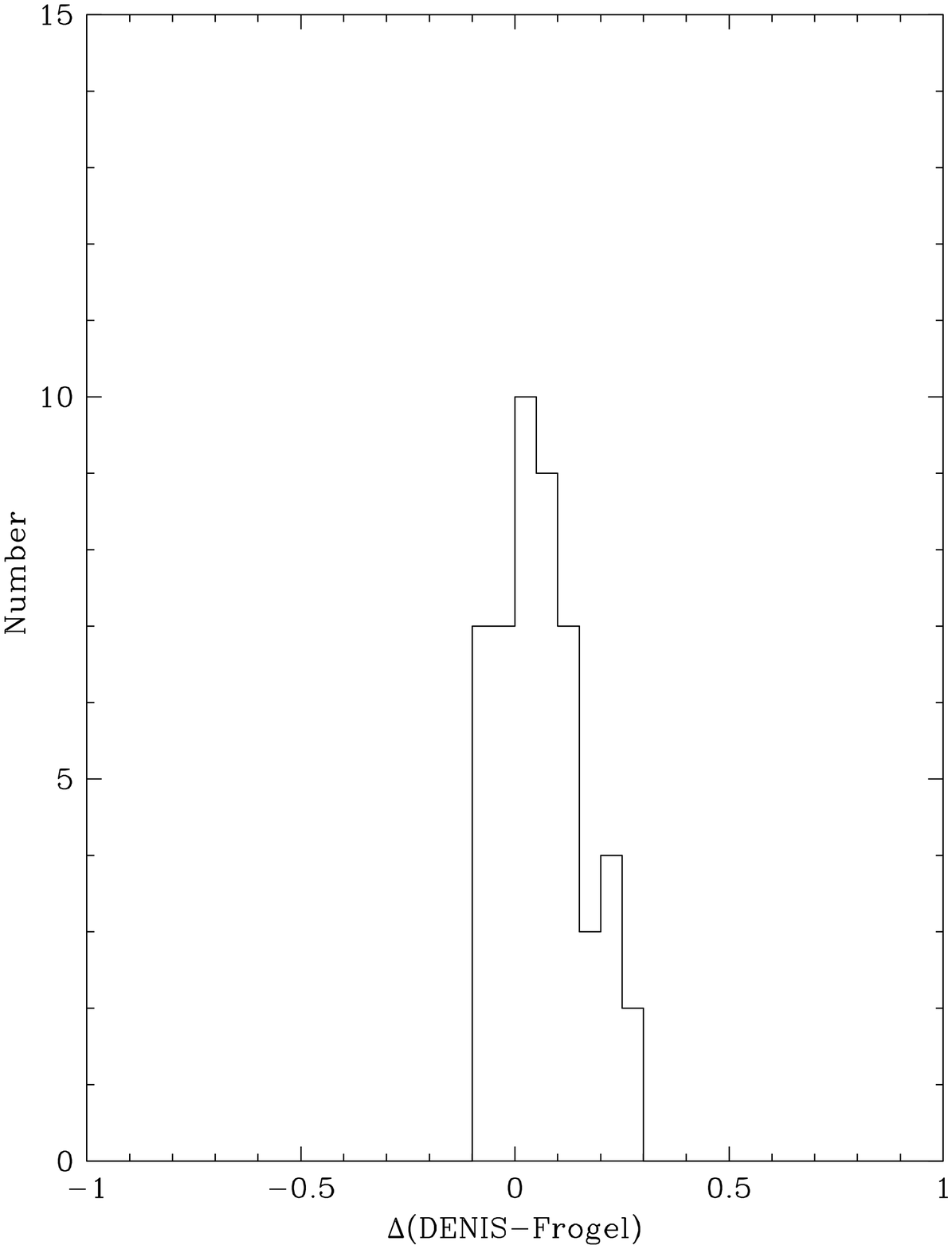}}

\caption{Histograms of differences between DENIS
and FW data for ISOGAL/MACHO objects in common in NGC\,6522. The left histogram
represents the $I$ data; $<\Delta I>$ = --0.002, s.d = 0.4. The centre
histogram represents the $J$ data; $<\Delta J>$ = 0.24 s.d. 0.12. The
right histogram represents $K$; $<\Delta K>$ = 0.04 s.d. 0.11.}

\end{figure*}

Fig 1 shows histograms of differences between DENIS and 2MASS $J$ and $K_S$
data for MACHO variables identified in both data sets. It is seen that the
agreement is satisfactory, both $\Delta J$ and $\Delta K_S$ being less than
0.01 mag on average, with $\sigma$ $\sim$ 0.12 and 0.09 for the two bands
respectively. The two surveys were conducted during the same year and
season. We have also plotted the differences between the DENIS and 2MASS
photometry for all stars in common in the ISOGAL field against (J--K) and
find no significant trend.
 
Fig 2 shows histograms of the differences between the DENIS and the FW
data for NGC\,6522 in the sense (DENIS -- FW). It should be noted
that the size of the sample is small and some degree of variability is
likely in almost all the stars. The FW  $I$-band data are from the
photographic survey by Blanco et al. (1984) and Blanco (1986). As might be
expected, $\Delta I$ shows a large scatter (0.4 mag) but its average value
is almost zero (note that the number of objects compared is small).

$\Delta J$ shows an average of 0.24\,mag with $\sigma$ = 0.12. According to
the conventional transformation between the Caltech and SAAO (and similar)
systems, defined by observations of standard stars in common, part at least
of the difference ($\sim$0.14 mag) can be attributed to the effective
wavelengths of the filters. There is evidence that transformation of the $J$
mags of late-type variables requires a stronger colour term than the
standard stars (Glass 1993).

$\Delta K$ is 0.04 with $\sigma$ = 0.11. Differences arise from the
shorter cut-on and cut-off wavelengths of the $K_S$ filter relative to $K$.
These will have strong effects if the spectra show well-developed H$_2$O or
CO bands as is the case for Mira variables and other very late-type M stars.
$K_S$ will include more of the water-vapour band and less of the CO first
overtone band than $K$.

\begin{figure} 
%fig3
\centering
\epsfxsize=6.5cm
\epsffile[20 17 592 779]{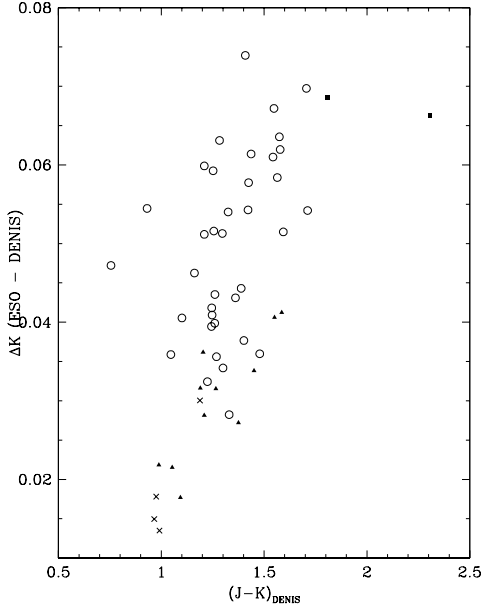}
\caption{Differences between ESO standard $K$ magnitudes and DENIS $K_S$
obtained by synthetic photometry, based on spectrophotometry by LW, for
Giants (crosses), SRVs (triangles), field Miras (open circles) and Bulge
Miras (solid squares).}
\end{figure}

\begin{figure*}
%fig4a
{\epsfxsize=4.3cm \epsffile[20 17 592 779]{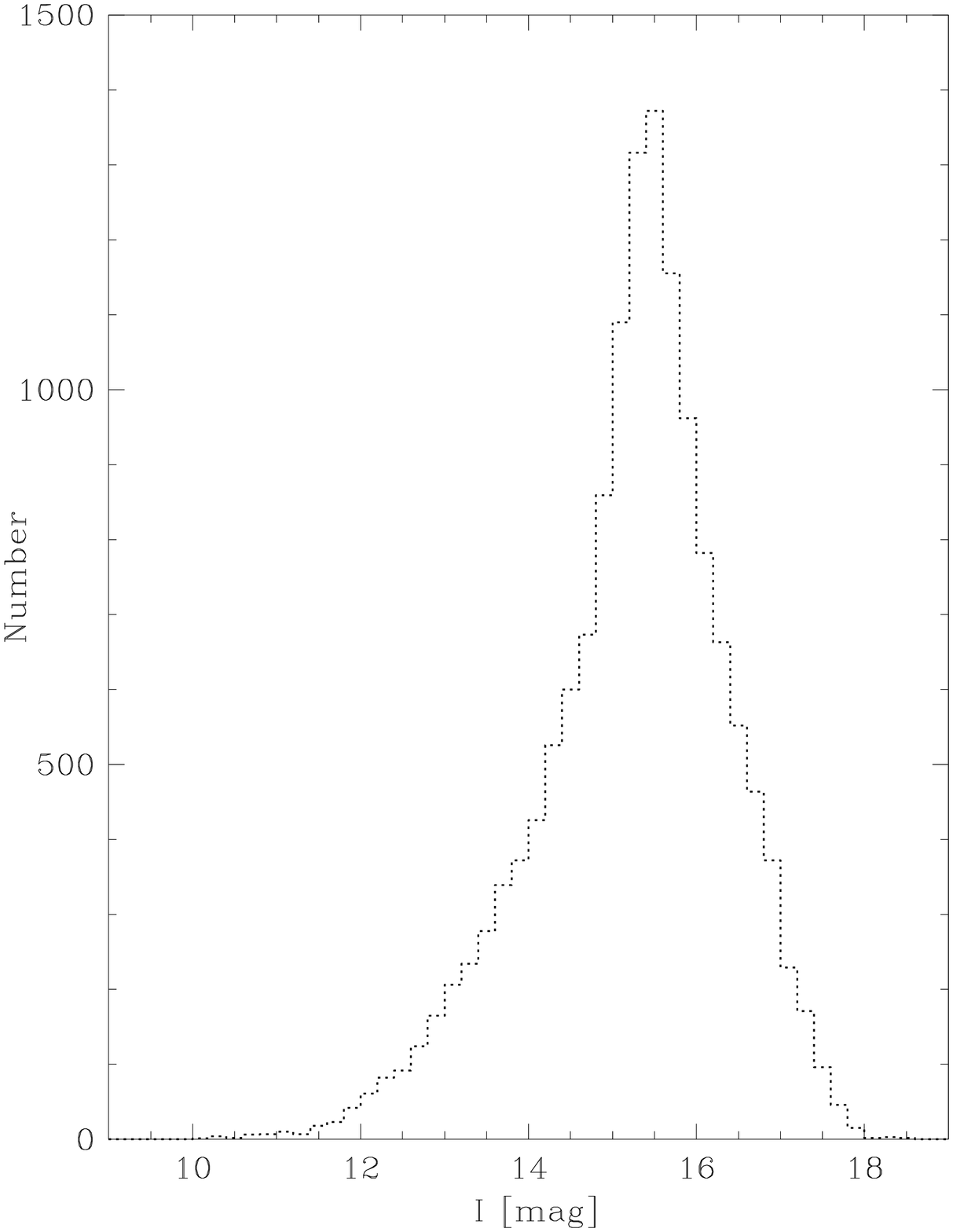}
\epsfxsize=4.3cm \epsffile[20 17 592 779]{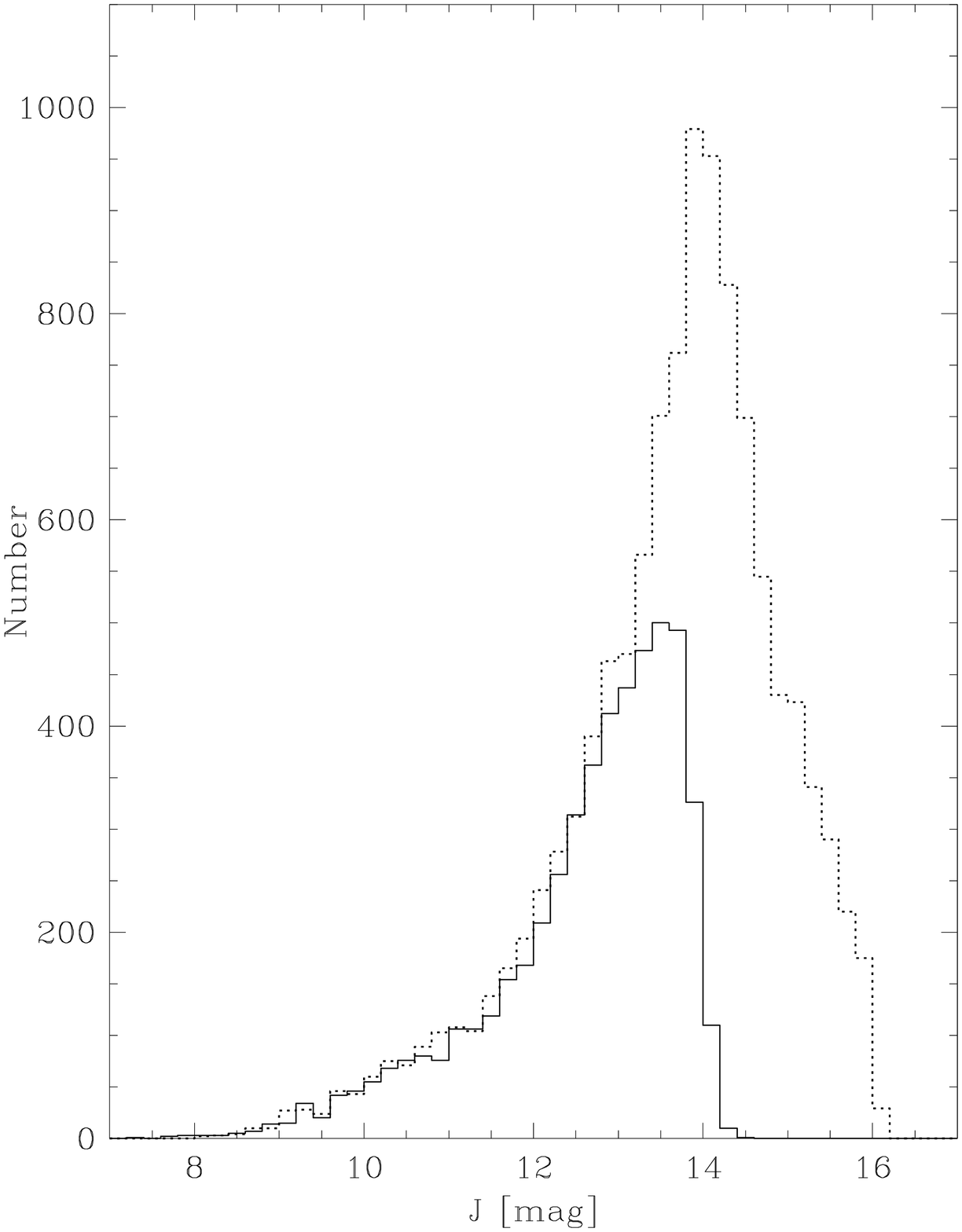}
\epsfxsize=4.3cm \epsffile[20 17 592 779]{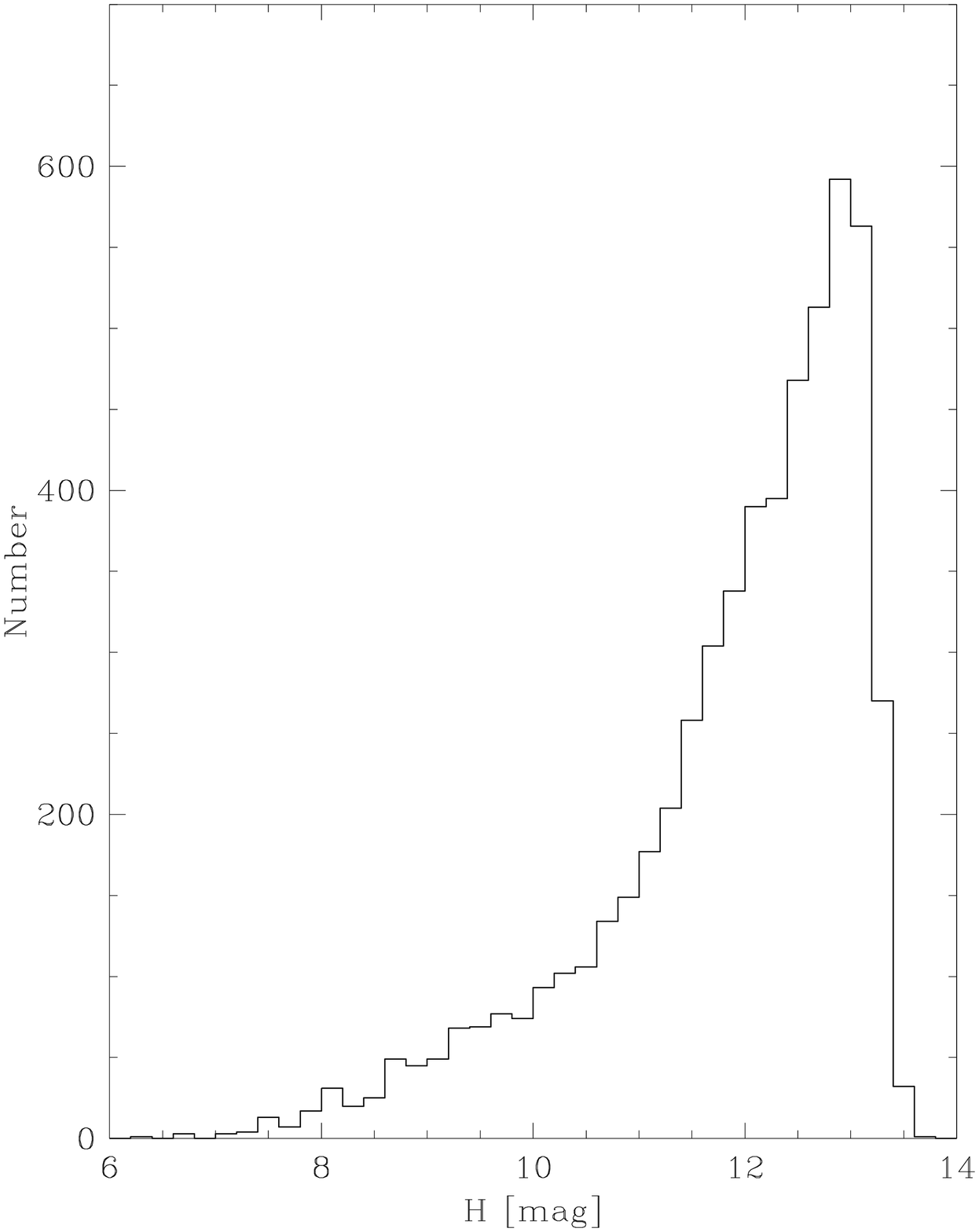}
\epsfxsize=4.3cm \epsffile[20 17 592 779]{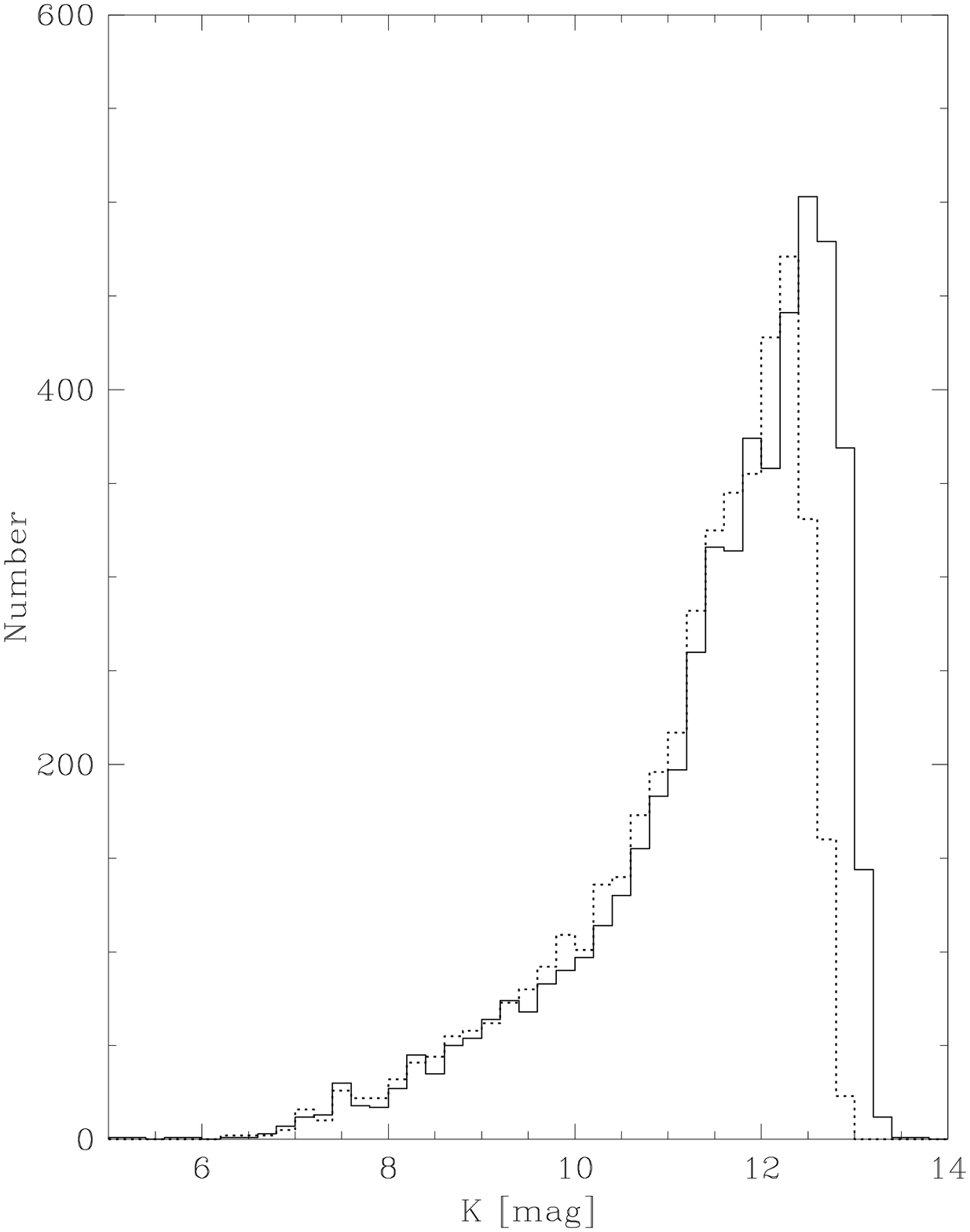}}

\caption{Histograms of near-infrared photometry of objects from the DENIS
and 2MASS surveys in SgrI. From left to right: $I$, $J$, $H$ and $K_S$. In the $J$
and $K_S$ diagrams, the DENIS histograms are shown dotted and the 2MASS
solid.}

\end{figure*}

We have constructed synthetic $K_{\rm ESO}$ and $K_{\rm DENIS}$ data from
the Lan\c{c}on \& Wood (2000, hereafter LW) spectrophotometry (see also
section 5). In this exercise the spectral energy distributions were
convolved with the DENIS filter profiles and the estimated average
atmospheric transmission at La Silla. The zero-points of the synthetic
photometry were obtained from spectrophotometry of Vega, which was taken to
have magnitude 0.0 at all wavelengths. We found that most categories of
late-type stars show $\Delta (K-K_{\rm DENIS})$ $\sim0.03$ mag with s.d.
$\pm$0.01 mag. However, the difference increases to 0.05 with p-p scatter
$\pm$0.025 for field Miras (see fig.\ 3). Two long-period Bulge Miras in the
LW sample show $\Delta (K-K_{\rm DENIS})$ $\sim$ 0.07 mag.

\begin{figure} 
%fig 5 
\epsfxsize=7.8cm 
\epsffile[20 17 592 779]{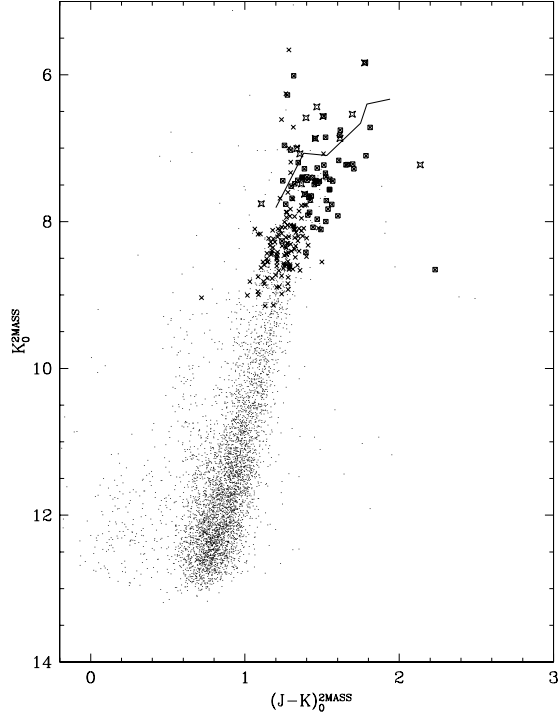}

\caption{$K_{S,0},(J-K_S)_0$ diagram from 2MASS data. The small dots are
stars from the general field of SgrI.  Crosses represent
ISOGAL/MACHO SRVs with low mass-loss; squares those with high. Miras
in the ISOGAL/MACHO field are shown with starred symbols. The
segmented line shows magnitudes and colours averaged by period groups from
Glass et al.\ (1995). The Miras are overlapped by the more luminous SRVs.
Note that, for M stars, $K$ is approximately related to $m_{\rm bol}$ by
$m_{\rm bol} \sim K+3$. See text concerning the isolated ISOGAL/MACHO star
at ($K_{S,0},(J-K_S)_0$) = (8.65, 2.23).}

\end{figure}

\section{Star counts and completeness}

Fig 4 shows  histograms of near-infrared photometry of objects from DENIS
and 2MASS in the Sgr\,I Baade's window and may be used to examine the
completeness of these surveys as functions of magnitudes in the
$J$ and $K_S$ bands. The area considered is slightly less than the total
ISOGAL field in order to avoid edge effects. 

At $J$ it is evident that there are slightly fewer 2MASS sources than DENIS
in most magnitude bins. The two histograms deviate systematically for $J$
$>$ 13.0 and indicate that DENIS is probably complete to about 13.5 ($\sim$
80\% of the maximum of the histogram).

In $K_S$ it is once again evident that the 2MASS detections are slightly
fewer than the DENIS to about 12.0. The limit of 2MASS appears to be about
0.2 mag fainter than DENIS.

DENIS $I$ appears complete to about mag 15.0 while 2MASS $H$ is complete to
about 12.5.

For both 2MASS and DENIS the sensitivity is mostly limited by confusion in
the J, H and $K_S$ bands. The densities of sources detected in DENIS reach 73
pixels/object at $I$, 6 pixels/object at $J$ and 15 pixels/object at $K_S$.
The corresponding figures for 2MASS are 13 pixels/object in $J$ and $K_S$
and 12 pixels/object in $H$. Almost certainly, the reliability of detection
and the accuracy of the photometry decreases towards the faint end, but a full
discussion of these issues awaits the publication of the catalogues.

There are 201 ISOGAL/MACHO stars in Sgr\,I, of which 196 were detected by
DENIS and 187 (of which 185 were in common with DENIS) by 2MASS. The two
2MASS stars not detected by DENIS lay along the edge of the field and may
have been lost due to slightly different astrometric solutions.

%sect5
\section{Colour-magnitude diagram}

\begin{figure}
%fig6 
\epsfxsize=8.3cm 
\epsffile[20 17 592 779]{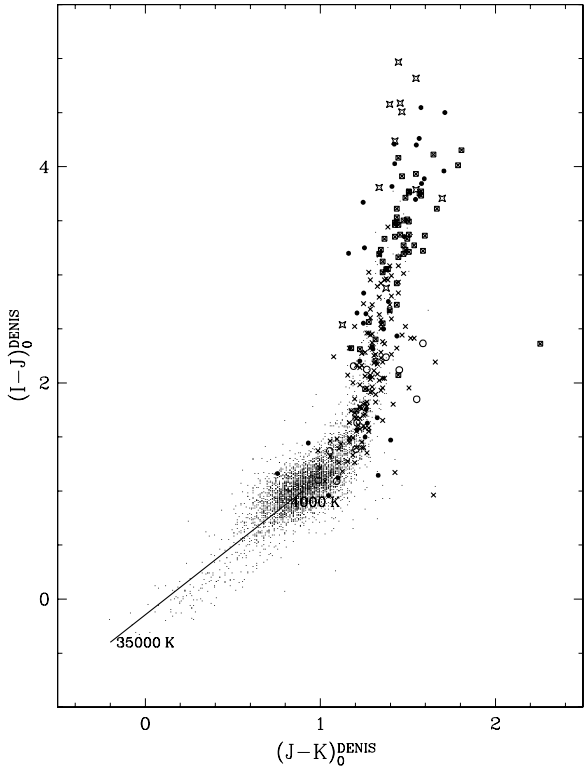}
\caption{$(I-J)_0$ vs $(J-K)_0$ for ISOGAL/MACHO from DENIS. Crosses represent
ISOGAL/MACHO SRVs with low mass-loss; squares those with high. Small points
represent the remainder of the DENIS detections in the SgrI field. The line
shows the location of earlier-type giants from Pickles (1998). Open circles
are synthetic colours of SRVs and closed circles of Miras derived from
the LW spectrophotometry.}
\end{figure}

Fig 5 shows the $K_{S,0},(J-K_S)_0$ diagram for the 2MASS data from Sgr I.
Although the 2MASS data show smaller scatter, the DENIS data show fewer
objects with outlying high $(J-K_S)$ colours. An investigation of such
objects from the general field of Sgr\,I showed that in DENIS some were not
recorded, some had $J$ but not $K$ detections and the remainder had ordinary
$J-K$ colours.  

We distinguish with special symbols the ISOGAL/MACHO variables with high
mass-loss, defined by [7] - [15] $>$ 0.6 and [15] $<$ 7.5, the latter
criterion being used to eliminate objects with high probable errors. The
diagram is dominated by the sequence of late-type giants having the
ISOGAL/MACHO objects towards the tip. The SRVs with infrared excesses are
generally redder and more luminous than the others. The Miras lie at or near
the tip and are overlapped by the SRVs. Some apparently luminous stars in
this diagram may lie in the foreground.

There is also a vertical sequence of objects at $(J-K_S)_0$ $\sim$ 0.6 which
may represent foreground stars.

The isolated star at ($K_{S,0},(J-K_S)_0$) = (8.65, 2.23) is the reddest
object in Fig 9 (the [15] vs [7]--[15] diagram) of Glass et al (1999). Because
of doubt about its (preliminary) ISOGAL photometry, it was omitted from
Table 3 of the same paper. It is listed as J175850.9--290106 in Alard et
al.\ (2000), with log $P$ = 2.158: (144\,d, uncertain). Its excess appears
to begin at least by the $K$-band. It is very red at mid-IR wavelengths,
with $K-[7]$ = 1.61 and [7] -- [15] = 1.98 (Omont et al.\ 2001).

In addition, during the preparation of this diagram, two ISOGAL/MACHO stars
(J175903.0--290137 and JJ175916.9--290225) that appeared to be about a
magnitude fainter than the others in $K_S$ were investigated and found to
have been dropped from the final ISOGAL Catalogue (Omont et al., 2001),
i.e., they are MACHO stars not identified with reliable mid-infrared
sources.

\section{Colour-colour diagrams}

In what follows we include some solar neighbourhood giants for comparison
purposes. These have been derived from the spectrophotometric atlases of
LW and Pickles (1998). Their colours have been
calculated assuming the ESO and DENIS photometric response functions.
Recourse to this `synthetic' photometry, based on spectrophotometry, is
necessary because no data that combine the $I$ band with $JHK_S$ exist
outside the DENIS work.

\begin{figure} 
%fig 7 
\epsfxsize=8.3cm 
\epsffile[20 17 592 779]{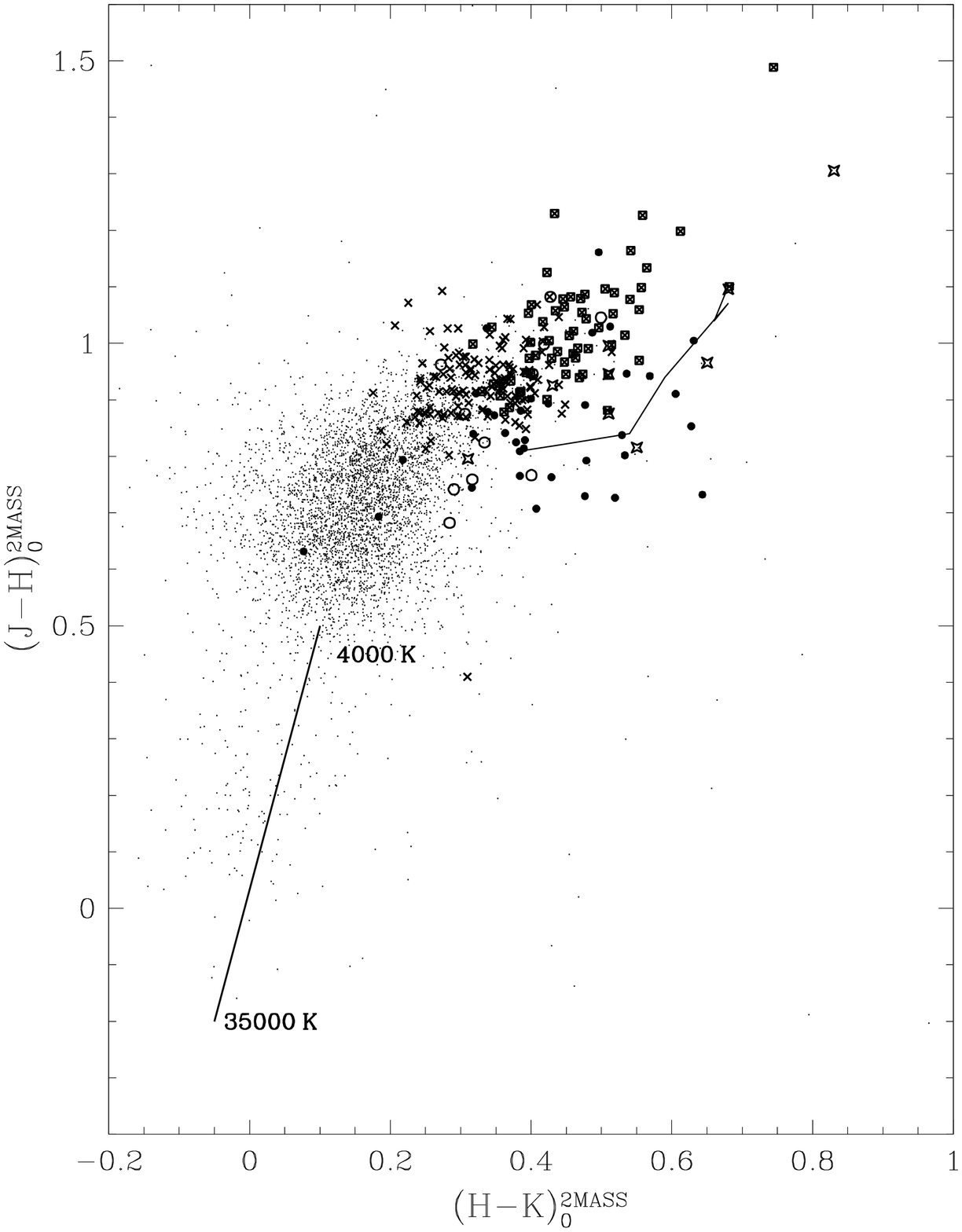}

\caption{$(J-H)_0,(H-K_S)_0$ diagram from 2MASS.  Crosses represent
ISOGAL/MACHO SRVs with low mass-loss; squares those with high. Small points
represent the remainder of stars in Sgr\,I. The straight line shows the
location of earlier-type giants from Pickles (1998). The bent line is the
location of the average colours of Mira variables in Sgr I, taken from Glass
et al. (1995) and shifted by 0.07 mag in $H-K$ to allow for the difference
between the SAAO and 2MASS $K$ filters. Open circles are synthetic colours
of SRVs and solid circles are Miras derived from the LW
spectrophotometry. The three bluest (in $H-K$) Mira points are amongst those
representing the 150d hot Mira S Car.}

\end{figure}

\begin{figure*} 
%fig 8
\epsfxsize=8.3cm

\centerline {\epsfbox[10 10 760 600]{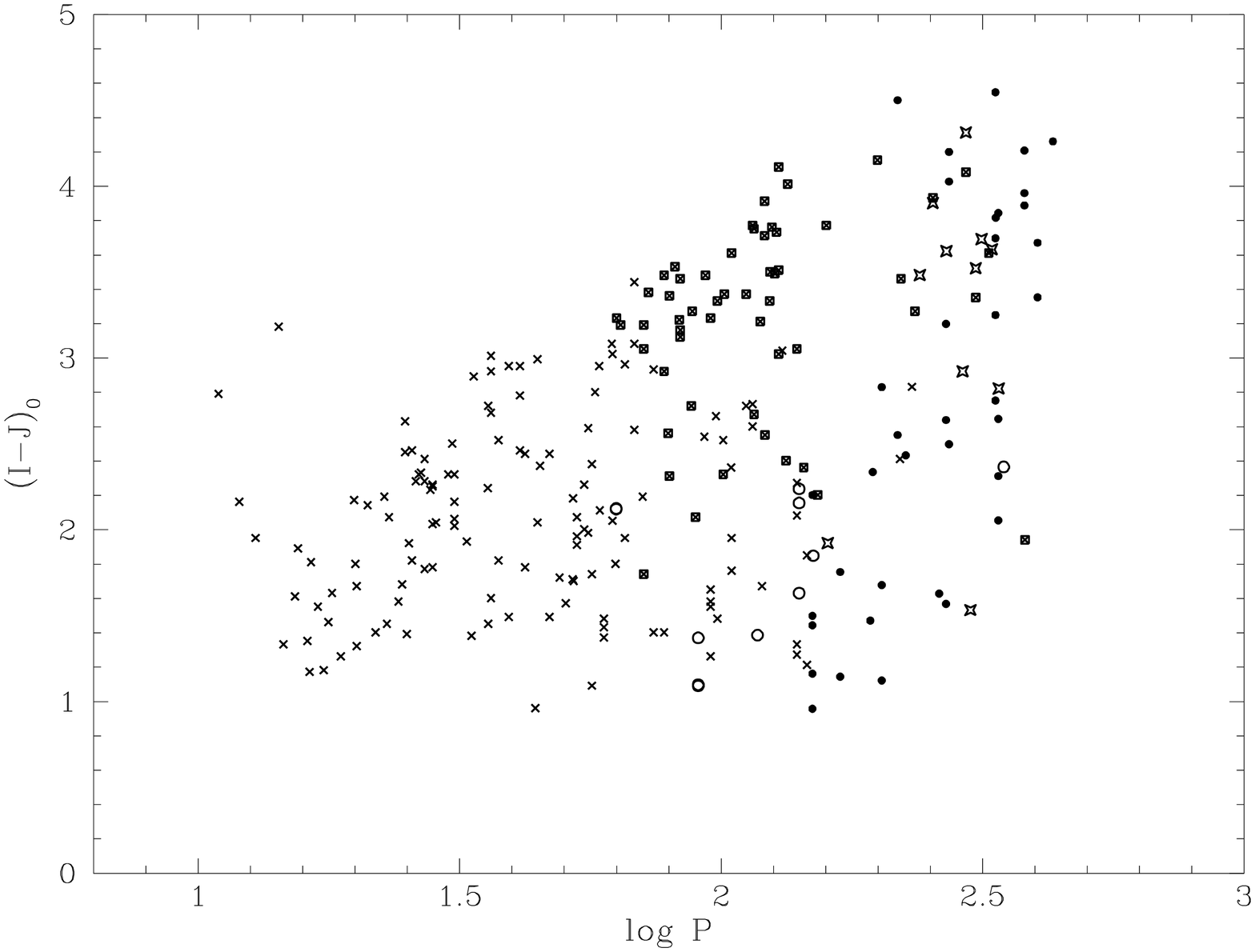} \epsfxsize=8.3cm \epsfbox[10 10 760
600]{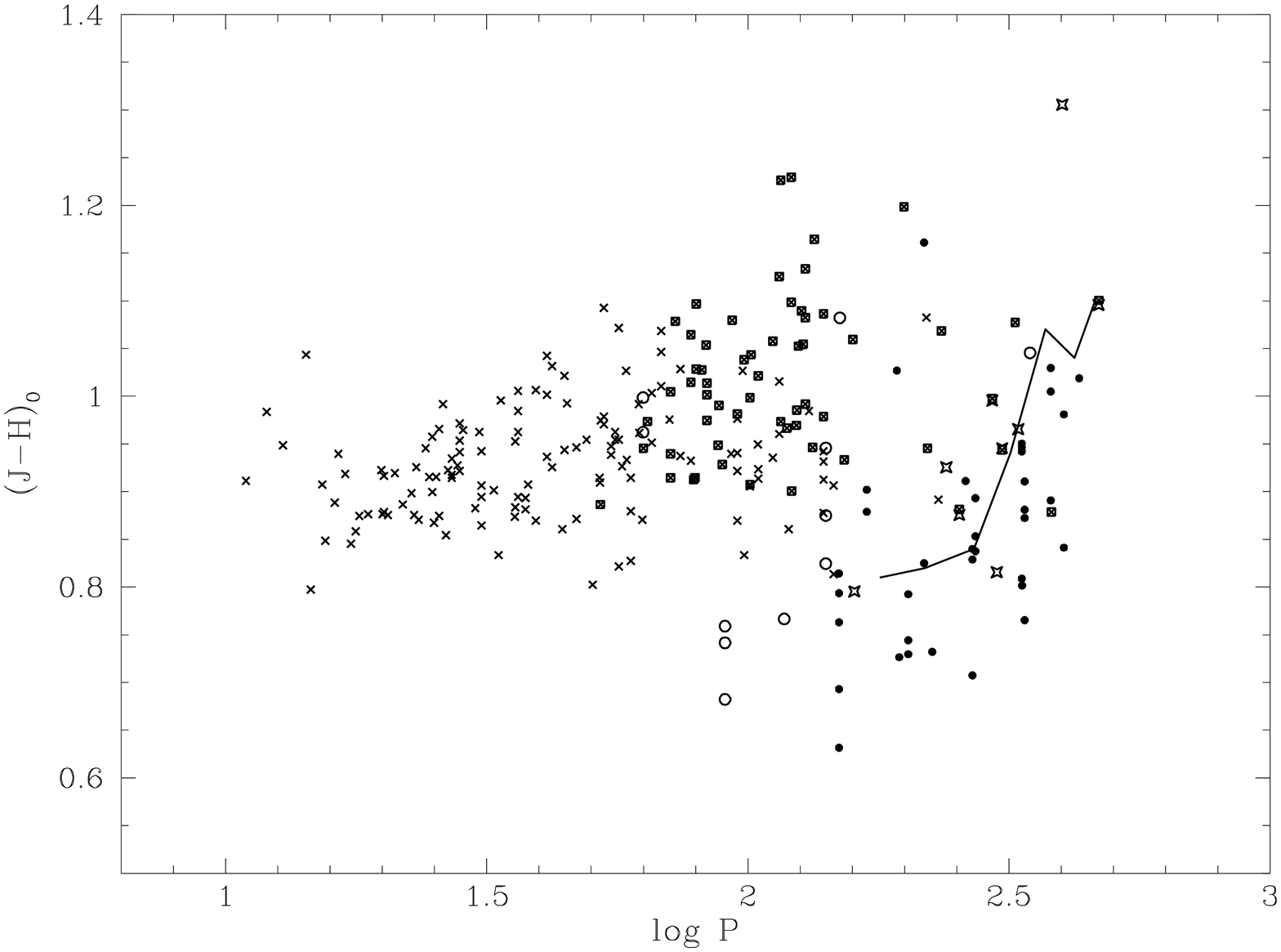}}

\epsfxsize=8.3cm
\centerline {\epsfbox[10 10 760 600]{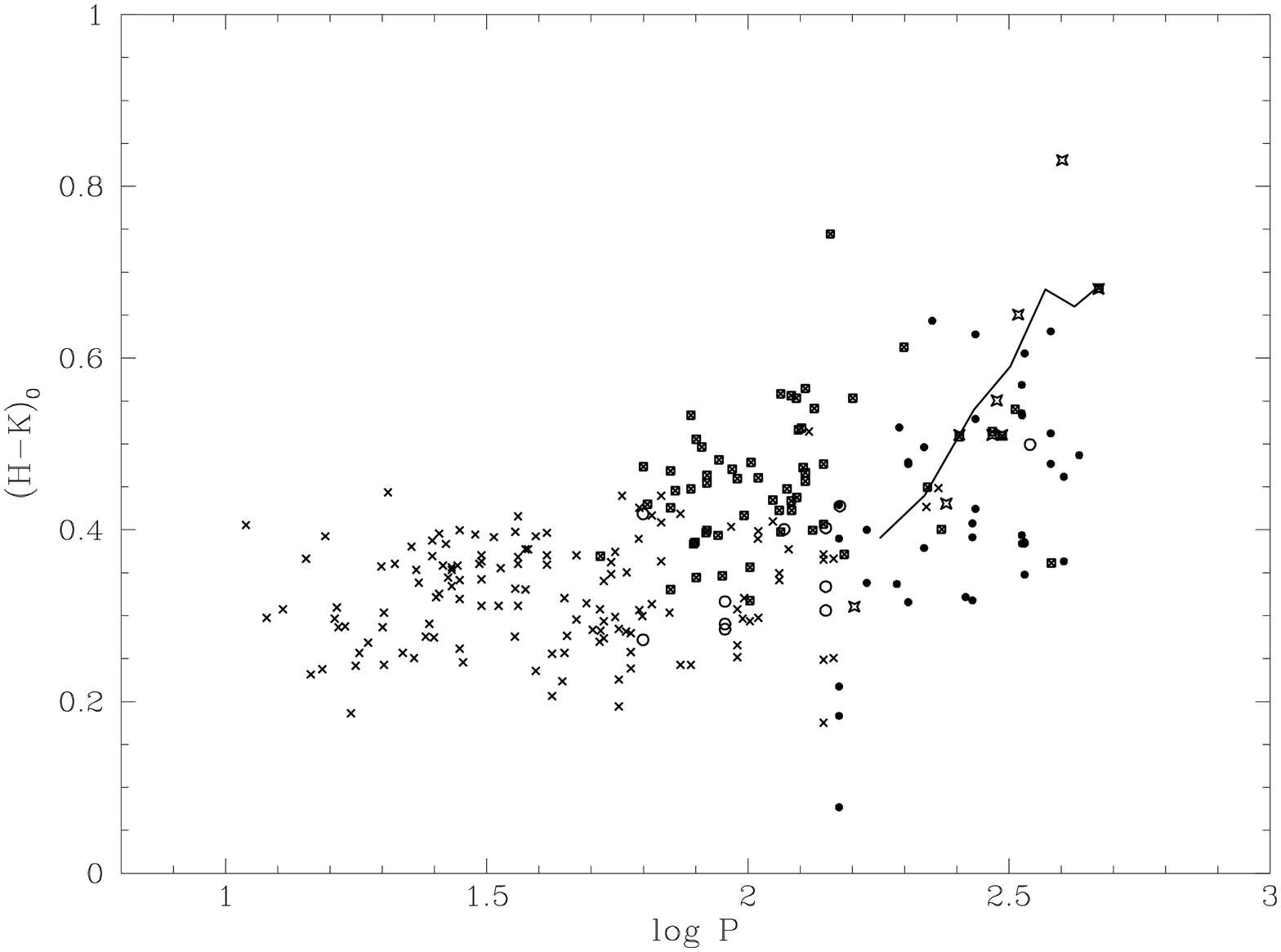} \epsfxsize=8.3cm \epsfbox[10 10 760
600]{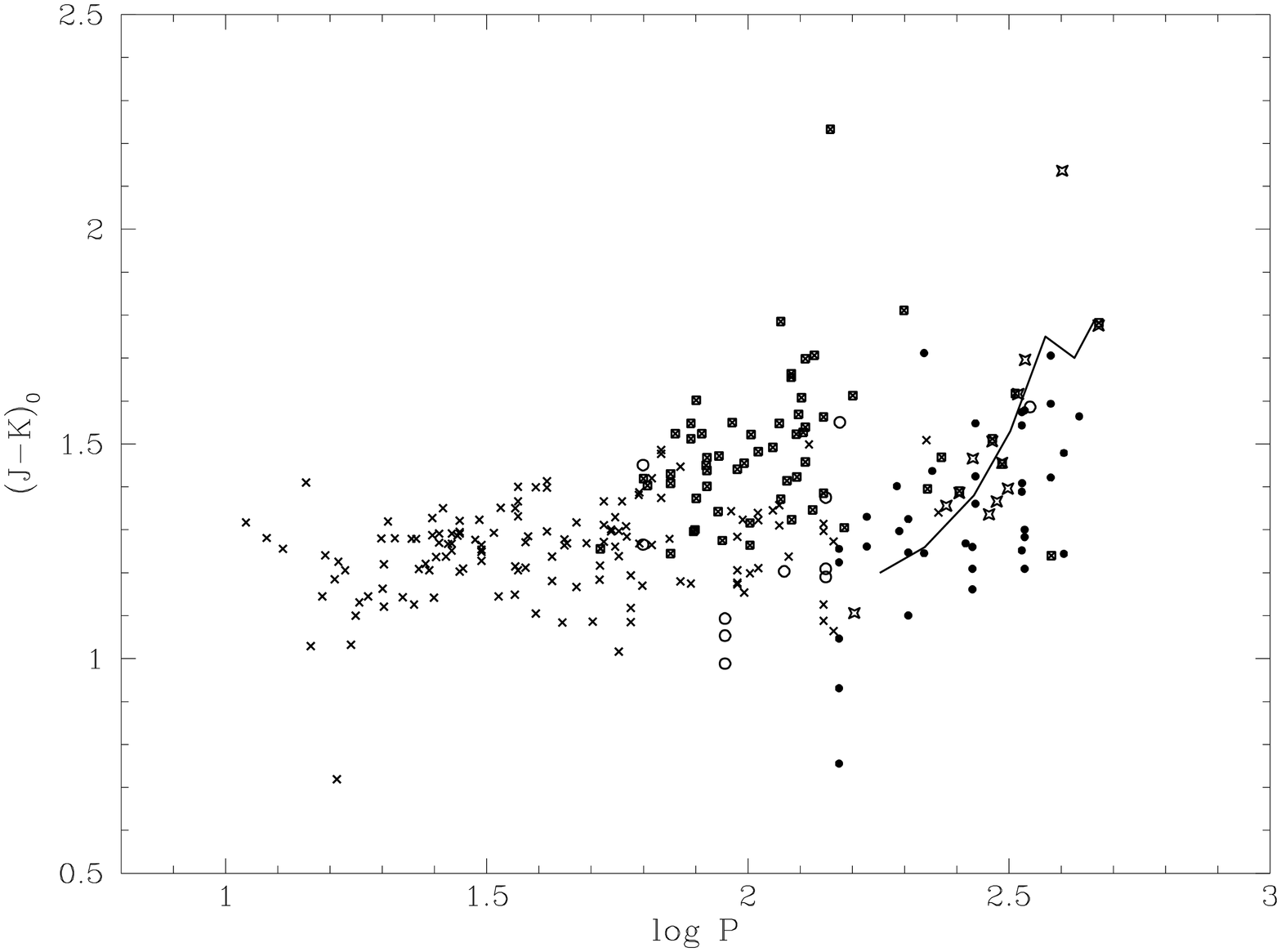}}

\caption{Colour-period diagrams for ISOGAL/MACHO stars. 
(a) (Top left) $(I-J)_0$ vs log $P$ from DENIS. The crosses represent SRVs
without appreciable mass-loss and solid squares those with mass-loss. The stared symbols
represent Mira variables from ISOGAL/MACHO. Also included are LW SRVs (open circles) and Miras (filled circles), calculated for
the DENIS filters. (b) (Top right) $(J-H)_0$ vs log $P$ from 2MASS for
ISOGAL/MACHO Bulge SRVs. The averaged colours of Sgr I Miras for various
period groups from Glass et al.\ (1995) are also shown. Other symbols as in
(a), above. The displacement to lower $(J-H)$ of the Miras is a well-known
phenomenon but its dependence on pulsation period is shown here for the
first time. (c) (Bottom left) $(H-K_S)_0$ vs log $P$ from 2MASS. Information
as in (b), above. (d) (Bottom right) $(J-K_S)_0$ vs log $P$ from 2MASS.
Information as in (b), above.}

\end{figure*}

Fig 6 presents the $(I-J)_0$ vs $(J-K_S)_0$ diagram. There is a continuous
well-defined sequence from left to right. The LW and
Pickles (1998) stars cover the giants from medium to late spectral types. 
By comparison with the colours of the latter, it is apparent that the
clumped DENIS objects are predominantly of K- and early M-type, with smaller
numbers of earlier and later types. The ISOGAL/MACHO stars without dust show
a rapid increase in $(I-J)_0$ with $(J-K_S)_0$ colour as the $I$-band
strength decreases due to increasing TiO and VO absorption at lower
temperatures. Finally, the most extreme colours are shown by the stars with
dust excesses that are also the most luminous. The sequence shows much
greater scatter about a given point in $(I-K_S)_0$ and $(I-J)_0$ than in
$(J-K_S)_0$.

Fig 7 shows the  $(J-H)_0,(H-K_S)_0$ diagram. Again a clear sequence is seen
from the LW and Pickles (1998) stars through the
great mass of 2MASS objects, to the ISOGAL/MACHO stars. The average colours
of Miras in Sgr I from Glass et al. (1995), computed for various period
groups and shifted to the right by 0.07 mag, to allow for the difference in
the $K$ filters, are also shown as a solid line.

In general, the $J-H$ and $H-K$ colours, which are sensitive to metallicity
(see comparison of LMC and Sgr\,I Miras in Glass et al. 1995) and
atmospheric extension (Bessell et al.\ 1989), of the SRVs and Miras of the
Baade's Window population agree closely with the solar neighbourhood sample.

\section{Colour-period diagrams}

The periods of the SRVs are by their very nature not very well-defined. The
techniques for determining them from the MACHO data and their limitations
have been described by Alard et al. (2001). Essentially, the periods given
here are those that were dominant during the years of the MACHO
observations. In some cases, other (usually nearby) periods had nearly equal
Fourier amplitudes.

Fig 8 shows the colour-period plots for the available samples. 

One should recollect that appreciable mass-loss is only seen for those stars
that have periods in excess of 70d (Alard et al. 2001), but a longer period
does not guarantee that it will occur. This is also true of field SRVs, as
found from CO observations (Kerschbaum Olofsson \& Hron 1996). The colours
of Miras can vary by 0.1 mag or more around a cycle. Note that an individual
variable star from LW may be represented by more than one point in these
diagrams since their observations were often made at more than one phase.

\subsection{$I-J$ vs log $P$}

In all colours it is apparent that the colours of the SRVs increase on the
average with period. However, it is noticeable that the scatter in $I$ at a
given period is 3--4 mags whereas in the other colours it is much less (up
to 0.7). Alvarez et al. (2000) show that there is a close relationship (s.d.
= 0.09 mag) of wide validity between $I-M_{\rm bol}$ and $(I-J)$ colour for
O-rich stars. The Miras do not show any clear colour-period relation.

\subsection{$J-H$ vs log $P$}

Because of the poor overlap between the available 2MASS data and the
NGC\,6522 ISOGAL field, most of the stars in the diagrams involving the
$H$-band are from Sgr\,I. It is expected that the observed $(J-H)$ colours
will not differ significantly between the 2MASS and ESO or SAAO systems.

The semiregulars show a moderate increase in $(J-H)_0$ colour with period. 
They overlap, particularly at the longer-period end, the field SRVs (e.g.,
Kerschbaum \& Hron 1994, not shown in the figures in order to avoid
confusion). Our knowledge of the statistics of short-period field SRVs is
limited at present by their small amplitudes and the consequent difficulty
of detecting them by traditional methods. Only recently have systematic
studies become possible using photometry from the Hipparcos satellite
(Bedding \& Zijlstra 1998; Koen \& Laney, 2000).

That the Miras are displaced from the region of SRVs in the $J-H$, $H-K$
diagram has been known for many years (see e.g. Feast et al, 1982, and
references therein). The LW atlas shows definitively that stars which
exhibit strong H$_2$O absorption bands during their pulsation cycles have
large amplitudes. Now that this effect can be displayed on a $J-H$, log\,$P$
diagram (fig.\ 8b), the displacement caused by the water-vapour absorption
bands is seen to be quite dramatic at periods of about 150--200 days. In
this connection, it is also interesting to note that that the spectra of
Bulge Miras can be matched very closely by non-Mira spectra differing only
by the absence of H$_2$O features (LW).

The observed $J-H$ colours from Glass et al. (1995) are slightly redder on
average than the points derived from the LW spectrophotometry of field
Miras, though similar to the colours obtained by Kerschbaum (unpublished)
and others for field Miras. This difference may reflect the difficulty of
allowing for the terrestrial atmospheric water vapour absorption in deriving
the synthetic photometry.

We also comment that the LW points representing T Cen,
classed as a SRV (open circles) with period 90d (log $P$ = 1.95), are 
unusually low in this and the other colour-period diagrams. Its visual
amplitude of 3.5\,mag is excessive for a SRV, suggesting that it is a Mira
variable of unusually short period.

\subsection{$H-K_S$ vs log $P$}

The $(H-K_S)_0$ colours of the SRVs increase with period in much the same
way as the $(J-H)_0$.  It is more noticeable in this diagram than in the
$(J-H)$ vs log $P$ that the SRVs with [7] --[15] excesses are redder than
those without, at a given period. The LW points for
field SRVs are in agreement with the SRVs of similar period with low
mass-loss.

The Miras form a well-defined sequence which, however, starts below the
average $(H-K_S)_0$ of the SRVs at the same period ($\sim$180d). The Sgr\,I
relation appears to straddle the LW field Miras
satisfactorily, though the scatter of the latter is rather large.

\subsection{$J-K_S$ vs log $P$}

The $(J-K_S)_0$ index, like the others, increases steadily with period for
the SRVs, especially those with strong mass-loss from [7] --[15] photometry.
The redder colours of the mass-losing objects are even more obvious. 

\section{$K_S$ vs log$P$ diagram}

A version of this diagram, based on transformed 7\,$\mu$m mags, was
presented by Alard et al. (2001). It is given here with $K_S$ mags directly
from 2MASS. The conclusions of Alard et al. remain unchanged.

Absolute $K$ magnitudes of SRVs with periods are available for only a few
other samples. They include nearby objects with parallaxes from the
Hipparcos Catalogue (Bedding \& Zijlstra, 1998) and those in the Large
Magellanic Cloud (Wood, 2000) with periods also derived from MACHO.
Whitelock (1986) showed that SRVs in low-metallicity galactic globular
clusters exhibit a P-L sequence that falls about 0.8 mag below that of
Bedding \& Zijlstra.

\begin{figure} 
\epsfxsize=8.3cm 
\epsffile[22 16 750 600]{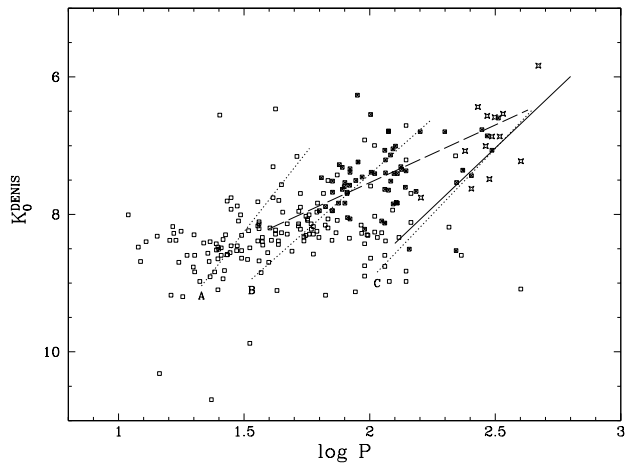}
\caption{$K_{S,0}$, log$P$ diagram for both Baade's Window fields. The
squares represent ISOGAL/MACHO sources; the filled ones have high mass-loss.
The solid line is the locus of Miras in Sgr I from Glass et al. (1995),
offset by 0.05 mag downwards in $K$ to allow for the difference in
photometric systems. Miras in the sample are indicated by the starred
symbols. The dashed line is the relationship suggested for local SRVs by
Bedding \& Zijlstra (1998). The dotted lines labelled A,B and C are eye fits
to LMC sequences by Wood (2000). We have taken 3.8 mags as the difference in
distance moduli between the LMC and the Bulge, as determined from the Mira
$P-L$ relation (Glass et al 1995), and we have assumed an LMC distance
modulus of 18.55.}
\end{figure}

In the Large Magellanic Cloud the Miras, the SRVs and the small-amplitude
SRVs seem to form distinct parallel sequences C, B, A etc (see fig.\ 9)
which have been identified by Wood (2000) as pulsators in the fundamental,
first and the next two higher overtones respectively. The present data do
not show such clearly separated sequences except perhaps between the Miras
and the others. The effect may be blurred in the Galactic Centre relative to
the Large Magellanic Cloud by the depth in the line of sight of the Bulge.
For example, the scatter of the {\it averaged} $K$ mags of Miras about the
P-L relation is 0.35 mag for the latter as compared to 0.13 for the LMC
(Glass et al. 1987).

Included in the diagram is a P-L sequence for the Hipparcos (nearby) SRVs
from Bedding and Zijlstra (1998). Similar lines, which cross the loci of the
various pulsation modes, were predicted by Vassiliadis \& Wood (1993) as
evolutionary sequences, their absolute luminosity depending on initial mass
and metallicity. The position of the line suggests that there is little
difference between the Bulge and local samples. The spectra of Mira
variables in the Bulge can also be matched well by local Miras (LW).

\section{Conclusions}

In the Baade's Windows $K_S$, $J-K$ diagram the most luminous stars are the
Miras and SRVs. The SRVs with significant mass-loss are generally more
luminous than those without.

The average $I-J$, $J-H$, $H-K$ and $J-K$ colours of the SRVs increase with
period but the effect is much less pronounced if the mass-losing stars are
omitted.  The scatter in the colours of SRVs is greatest in $I-J$. In all
the colours considered here, scatter increases with period.

The average $J-H$ colours of the shorter-period Bulge Miras ($\sim$ 200d)
are much bluer than the average of the SRVs with similar periods. A similar
but less conspicuous discontinuity occurs in the $H-K$ and $J-K$ colours.
This effect is attributed to the onset of H$_2$O absorption, which affects
$H$ and $K$ more than $J$.  In each of $J-H$, $H-K$ and $J-K$, the average
Mira colours show steeper rates of increase with period than the SRVs.

The colours of solar neighbourhood SRVs and Miras, derived for the DENIS and
2MASS systems from the LM spectrophotometry, agree with those of the Bulge.
In addition, the locations in the $K$, log\,$P$ diagram of the SRVs, which
should be sensitive to any differences in initial mass and metallicity, are
in good agreement.

\section{Acknowledgments}

We thank A. Lan\c{c}on for her help in calculating the synthetic near-IR
colours of stars from the spectrophotometry of LW and A. Omont for his
helpful comments. We also thank F. Kerschbaum for access to unpublished
photometry.

MS thanks SAAO and ISG thanks the IAP for their hospitality during visits
financed through the CNRS (France)/NRF (South Africa) agreement.

MS is supported by the Fonds zur F\"orderung der wissenschaftlichen
Forschung (FWF), Austria, under the project number J1971-PHY.

The DENIS project is supported, in France by the Institut National des
Sciences de l'Univers, the Education Ministry and the Centre National de la
Recherche Scientifique, in Germany by the State of Baden-W\"urtemberg, in
Spain by the DGICYT, in Italy by the Consiglio Nazionale delle Ricerche, in
Austria by the Fonds zur F\"orderung der wissenschaftlichen Forschung und
Bundesministerium f\"ur Wissenschaft und Forschung

This publication makes use of data products from the Two Micron All Sky
Survey, which is a joint prodect of the University of Massachusetts and the
Infrared Processing and Analysis Center/California Institute of Technology,
funded by the National Aeronautics and Space Administration and the National
Science Foundation.

\end{document}